# The Scale-free Network of Passwords : Visualization and Estimation of Empirical Passwords

Guo Xiujia, Chen Haibo, Liu Xuqin, Xu Xiangyu, Chen Zhong
Peking University, EECS
Beijing, China
guoxj@pku.edu.cn

***Abstract*—In this paper, we present a novel vision of large scale of empirical password sets available and improve the understanding of passwords by revealing their interconnections and considering the security on a level of the whole password set instead of one single password level. Through the visualization of Yahoo, Phpbb, 12306, etc. we, for the first time, show what the spatial structure of empirical password sets are like and take the community and clustering patterns of the passwords into account to shed lights on the definition of popularity of a password based on their frequency and degree separately. Furthermore, we propose a model of statistical guessing attack from the perspective of the data's topological space, which provide an explanation of the "cracking curve". We also give a lower bound of the minimum size of the dictionary needed to compromise arbitrary ratio of any given password set by proving that it is equivalent to the minimum dominating set problem, which is a NP-complete problem. Hence the minimal dictionary problem is also NP-complete.**

## I. INTRODUCTION

This Text password has been a ubiquitous way to access resources and services since 1960 and the attempts of password cracking have never stopped ever since. Especially in recent years, the leakage and compromise of large scale of password sets repeatedly remind us of the urgency of the security of password sets enhancement. Different password cracking metrics have been adopted in prior works and dictionary based password cracking still remain to be the most ubiquitous way in numerous attacks nowadays.

Conventional dictionary based password cracking metrics, such as the statistical guessing attack, usually start with a preprocessed dictionary and might involve some modification during the guessing process. Related research has been made by MSRA [1]. While due to the variance of original data set and dictionary size, the performance differs from one to the other. Bonneau made the first comparison in [2]. Nevertheless, dictionary based cracking metrics were proved effective and feasible in practice.

According to Bonneau, dictionary based cracking metrics was first proposed by Morris and Thompson in their seminal 1979 analysis of 3,000 passwords [3], and the two approaches, password cracking and semantic evaluation, was widely used ever since, even after Markov and PCFG were introduced. Though dictionary based cracking metrics have distinguished itself with feasible performance in practice and play a role of benchmark of the performance of a cracking technics, the reason why it works well remains UNKNOWN. Even though a lot of work have been made on password creation policy and password strength meters, the gap between our understanding of the security of one single password and the security of a whole password set was rarely discussed.

To prevent a password from being compromised, prior work have focused on improving the strength of a single password and block out passwords whose frequency exceed a particular threshold, which is intuitively reasonable but far from perfect.

For the former approach of assuring security, the first question is the definition of strong password, i.e. how to measure the security of a password and how to decide whether a password is strong or weak. Bonneau made a survey of the literature and propose the concept of Guessing Entropy, α-guess-work [3]. Common practice is the requirement of the length and the variety of characters in a password, such as at least 8 characters, have at least one lower case character, have at least one capital letter, have at least one number, etc.

For the latter approach of maintaining a blacklist of common passwords, it seems to be a game of cat and mouse. For every password that is blocked, the user almost always make a way out by performing a minor modification on it, for instance, by adding some characters at the rear, changing one or two digits, switch the first character into upper case, or simply use some other weak password that is not included in the list. The minor modification not only make the blacklist useless, but also leave a potential threat to the whole system. For the same blacklist, if everyone makes his or her own minor modification based on some limited popular passwords, the results could be different but similar. For example, if we all submit "password" as our password and it was blocked, the possible choices after minor modification might be "password1", "password12", "password123", "p@assword", "Password", etc. As we will discuss in this paper, the leakage of one single vulnerable password could lead the compromise of password one after another, thus creating a link reaction and endanger a larger scale of accounts.

Our first contribution is the visualization of several empirical password sets including Yahoo!, phpbb, myspace,

honeynet, hotmail, 12306, and build a network based on the inter connection of the passwords. To our knowledge, this is the first visualization of large scale password sets in the form of networks.

The second contribution is the exploration of the spatial structure of empirical data of passwords and prove that the distribution of passwords is a scale-free network, which provide an explanation of the attacking curve that has long been observed in decades.

Our final contribution is the model of statistical guessing attack, which explain the “cracking curve” that has been observed in variety of cracking results. Based on the model, we also have a discussion on the popularity of passwords from the perspective of frequency and degree separately. At last, we focus on the optimal cracking problem which aims at cracking a password set with the minimal size dictionary needed. We prove that the optimal cracking problem is equal to one of the 24 classic NP-complete problems, the dominating set problem, and thus it is also a NP-complete problem.

## II. BACKGROUND

### A. The definition of distance between passwords

Since we are going to figure out the relations of passwords, the first thing we need to do is to define the relationship between two passwords. When we are going to do something, we estimate the Pros and Cons. Similarly, when we are going to know something, we list the similarity and difference. Hence, what is the difference of two passwords? How to measure the degree of similarity and difference? Passwords are strings of alphabets, numbers and special characters, so the first choice is how we measure the similarity and difference of two strings ------ the edit distance.

Passwords are strings of alphabets, numbers, and special characters. In this paper, we adopt edit distance for the measurement of passwords.

Edit distance is a way of quantifying how dissimilar two strings (e.g., words) are to one another by counting the minimum number of operations required to transform one string into the other. One of the simplest sets of edit operations is that defined by Levenshtein in 1966 [4].

Several definitions of edit distance were defined by using different sets of unit string operations. In this paper we use one of the most common and widely used variants called Levenshtein distance, which was named after Vladimir Levenshtein [5]. Levenshtein distance may also simply be referred to as "edit distance", although several variants exist [6].

### B. Computation and the Algorithm

Using Levenshtein's original operations, the edit distance was defined by the recurrence.

In 1964, Damerau published the first recursive version of algorithm which takes exponential time for computing the edit distance between two strings [7]. And it is commonly credited to Wagner and Fischer who proposed an improved version of dynamic programming algorithm that has both the time and space complexity of O(mn) [8].

## III. VISUALIZATION OF THE EMPIRICAL PASSWORD SETS

### A. prior method of analysing a password set

- Statistical Analyze

  In most of the case, the presentation of the passwords is a list of the statistical information of the passwords, like the frequency, i.e. the number of the passwords that is identical to itself. Some of them might include the user name, the email, the age, gender, the register time, and even credit card information. This is nothing but a list of the account information.

- Word cloud of password set

  Word cloud is another option when visualizing words. According to the homepage introduction of Wordle, which is an online word cloud service provider (Certain parts of Wordle are © IBM Corporation, The text and design of the web site itself are Copyright © 2008 Jonathan Feinberg, and all rights are reserved.), the word clouds generated from given text give greater prominence to words that have higher frequency in the source text. Note that the fonts, layouts and color schemes can be tweaked by the users.

  In [9], Wordle was set up to show features in the password set of Rockyou, such as the mixed numeric and text dates.

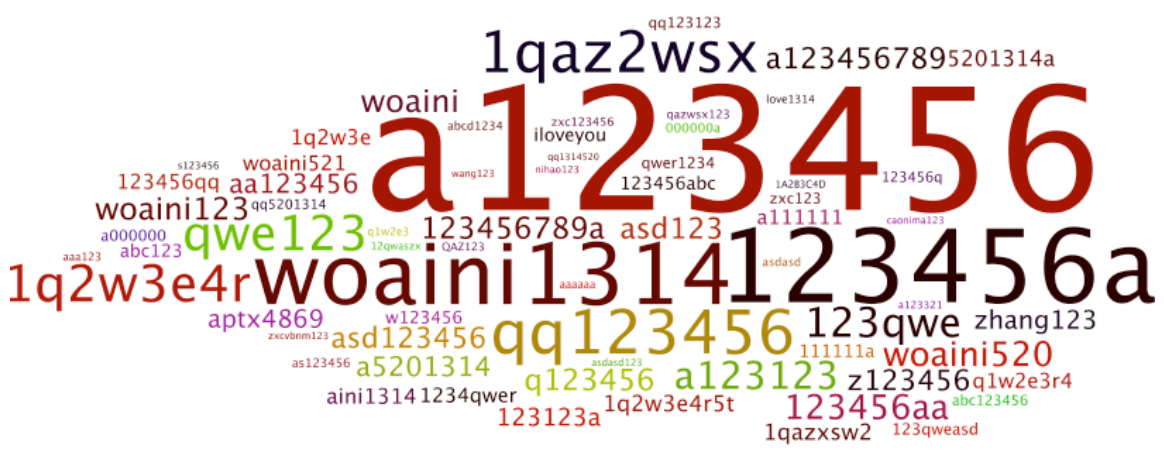

**Figure 1**

This method gives more straightforward and obvious information about the password set than the statistical table. Through the difference in size, the more important password distinguishes themselves from the ones that weights less. The variance in color also make the whole set more friendly than a simple dumb list of numbers. Furthermore, some patterns and features draw the attention of viewers. Sequences like “123456”, “qwer”( which is part of the first line of a keyboard),”123”,”woaini”(which means “I love you” in Chinese) appear in many of the passwords.

*B. Method of Our Visualization*

- The procedure of building a graph of the given data set:

  1) each unique password is a single node in the graph;

  2) two nodes are linked if the distance between two corresponding passwords is less than D(i,j) ( in this case, we assume it as 3);

  3) repeat step 2 until every two passwords in the data set have been compared;

  4) reorganize the graph and output the result.

  In practice, we compute the distance between every two passwords and store in form of adjacent table. In order to save space, we do not consider the form of adjacent matrix, which is O(n^2) in space complexity (n is the number of the total passwords in the date set).

- 1) The computing complexity analyze

  In order to analyze the computing complexity in password transformation, we need to make sure the basic components of passwords. For the characters used in passwords, we have the following chart:

| Type | numbers |
|---|---|
| Alphabet[a-zA-Z] | 26 |
| Digit[0-9] | 10 |
| Special Characters | ? |

The table above is slightly different from Weir's Listing of different string types in his paper: 1) the upper case of alphabet are taken into account; 2) Weir list 28 special characters (! @#$%^&*()-_=+ [] {};':",. <>/? ) in the table, which is part of the special characters on the keyboard. While some of the characters (such as blank space ~`, and some other special characters on mobile keyboard) are absent in Weir's list, we assume that Weir learn the result form the statistical work of empirical password set. It is worth noting that some special characters, like blank space, do exist in a large amount in some of our data set.

In theory, every item in the 128 ASCII（American Standard Code for Information Interchange）characters or the set of bytes [0-255] can be used in passwords under specific password policy. Hence the definition of symbols in our following analyze are as follows:

N: the number of different possible characters;

L(p): the length of password p;

D(pa,pb): the edit distance between password pa an pb;

Sp(k): the set of candidate passwords that has a distance k from password p, i.e. Sp(k) = { passwd | D(passwd,p)= k }.

Intuitively, Sp(0)={p};

# Sp(k): the number of passwords in set Sp(k);

When k=0,

#Sp(0)=1;

When k=1, i.e. the edit distance is 1.

L(p)+1:

Insertion: $C_{L(p)+1}^{1} * N$

L(p):

Substitution: $C_{L(p)}^{1} * (N\text{-}1)$

L(p)-1:

Deletion: $C_{L(P)}^{1}$

Thus

$$\#Sp(1)= C_{L(p)+1}^{1} * N + C_{L(P)}^{1} + C_{L(p)}^{1} * (N\text{-}1) = (2L(p)+1) * N$$

When k=2, i.e. the edit distance is 2.

L(p)+2:

Insertion twice: $C_{L(p)+2}^{2} * N^{2}$

L(p)+1:

Substitution and insertion: $C_{L(p)}^{1} * (N\text{-}1) * C_{L(p)+1}^{1} * N$

L(p):

Insertion and Deletion: $C_{L(P)}^{1} * C_{L(P)}^{1} * N$

Substitution twice: $C_{L(p)}^{2} * (N\text{-}1)^{2}$

L(p)-1:

Deletion and Substitution: $C_{L(p)}^{1} * C_{L(p)\text{-}1}^{1} * (N\text{-}1)$

L(p)-2:

Deletion twice: $C_{L(p)}^{2}$

Sum up all the situations and

$$\#Sp(2) = \left(\frac{3}{2}L^{2}(p) + \frac{3}{2}L(p) + 1\right) N^{2} \text{-} L(p) \cdot N$$

It can be proved that the number of passwords

$$\# Sp(k) = O(L^{k}(p) \cdot N^{k})$$

Thus the practically important case is k=1,2, and 3 [10].

*C. Examples of Our Visualization*

In order to compare with previous practice of visualization method, we will also take the top 100 password of the 12306 leakage as the example. The the adjacent table is taken as input for Gephi and the output is the graph of the network within the distance of 1, 2 and 3 separately. As shown in the picture, the visualization of 12306's top 100 passwords is as follows:

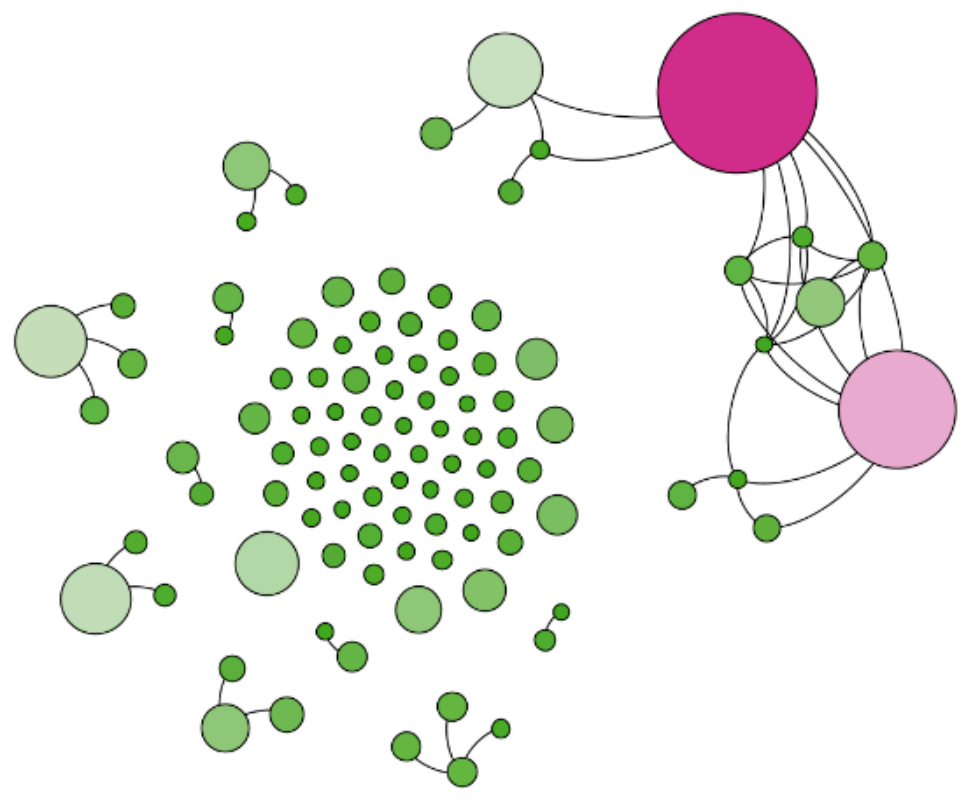

**Figure 2**

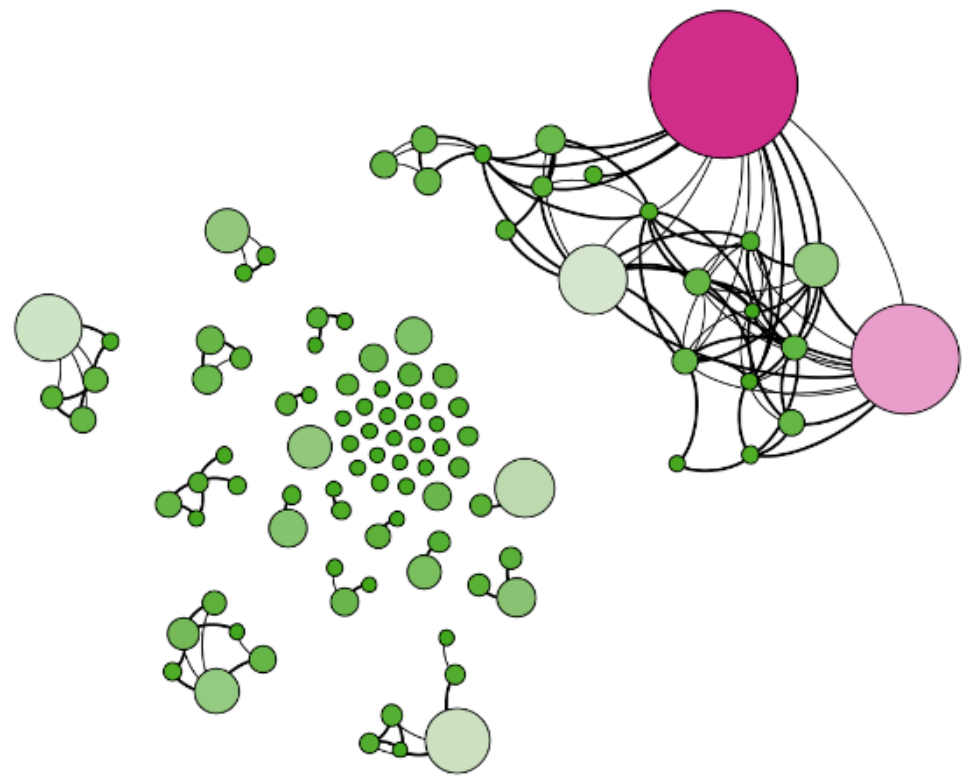

**Figure 3**

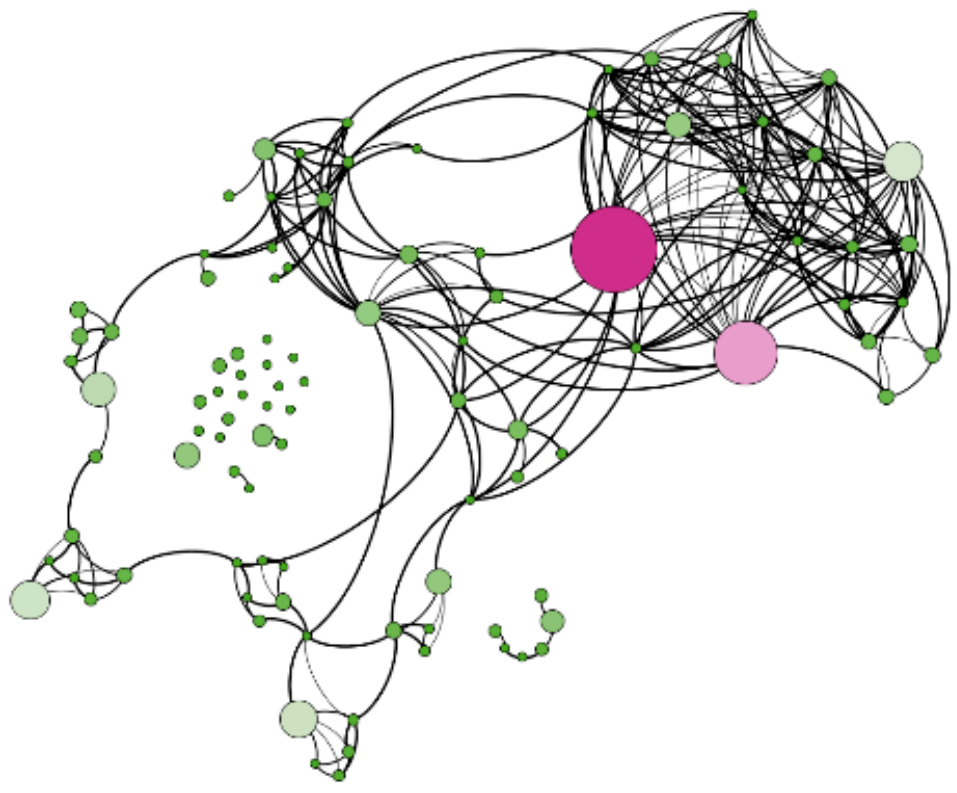

**Figure 4**

Figure 2 is the graph of the top 100 passwords within edit distance 1 in 12306 and figure 3 and figure 4 is the graph within 2 and 3 separately. As we can see in the graph, the number of edges increase as the edit distance grows. The whole graph is a network of passwords with some isolated nodes.

For brevity, we also only present the nodes that has a connection to others and show the password as their own label in figure 5 to make it more clear and straightforward.

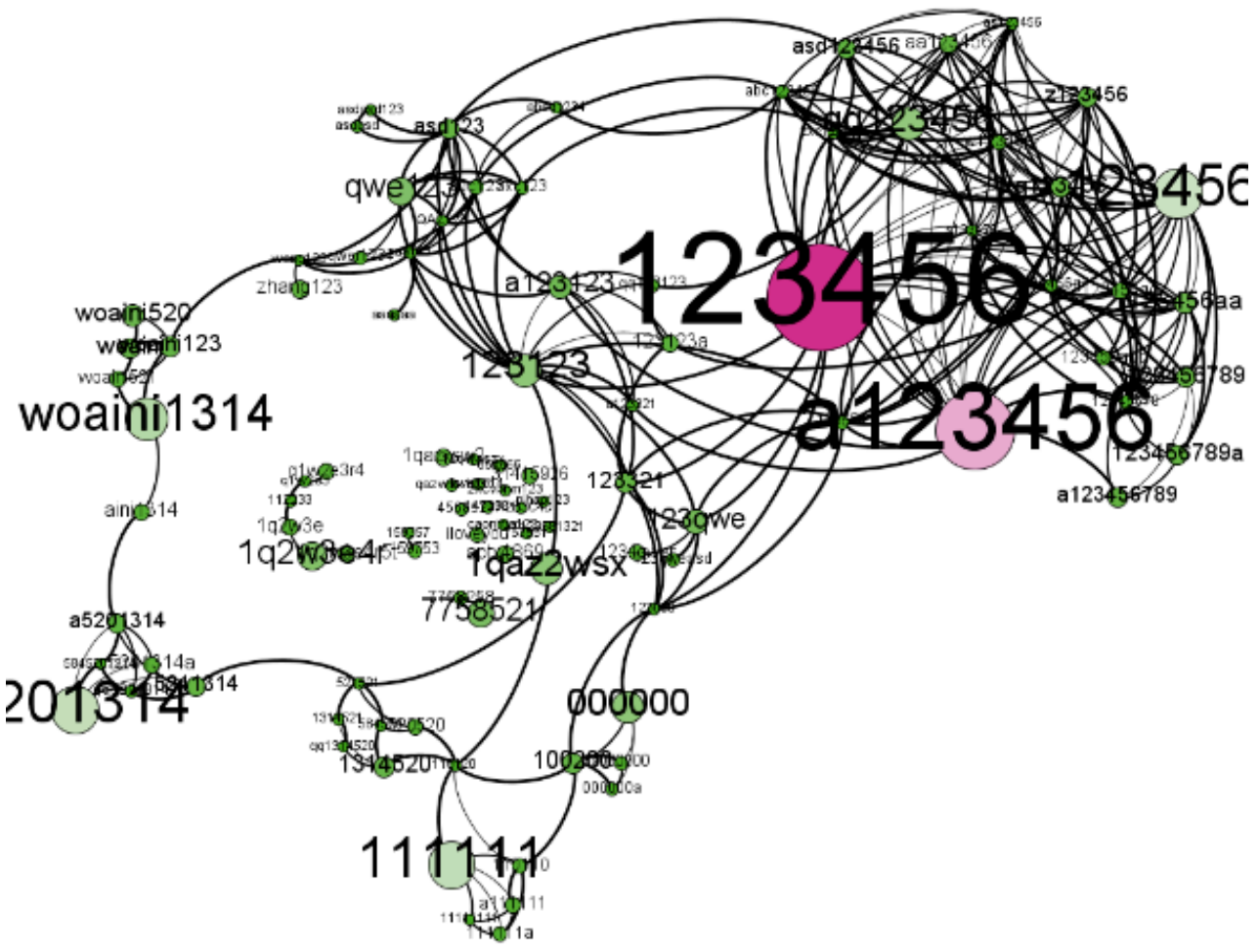

**Figure 5**

From the example of the top 100 passwords of the 12306, we expose the evolution of the password network within the distance k = 1, 2, and 3 and visualize the distribution of a empirical password set.

To further analyze the distribution of the data, we take the community and clustering method to separate the network apart and give a more clear vision of the structure of the data set. Again, we take the top 100 12306 password set for the example in figure 6.

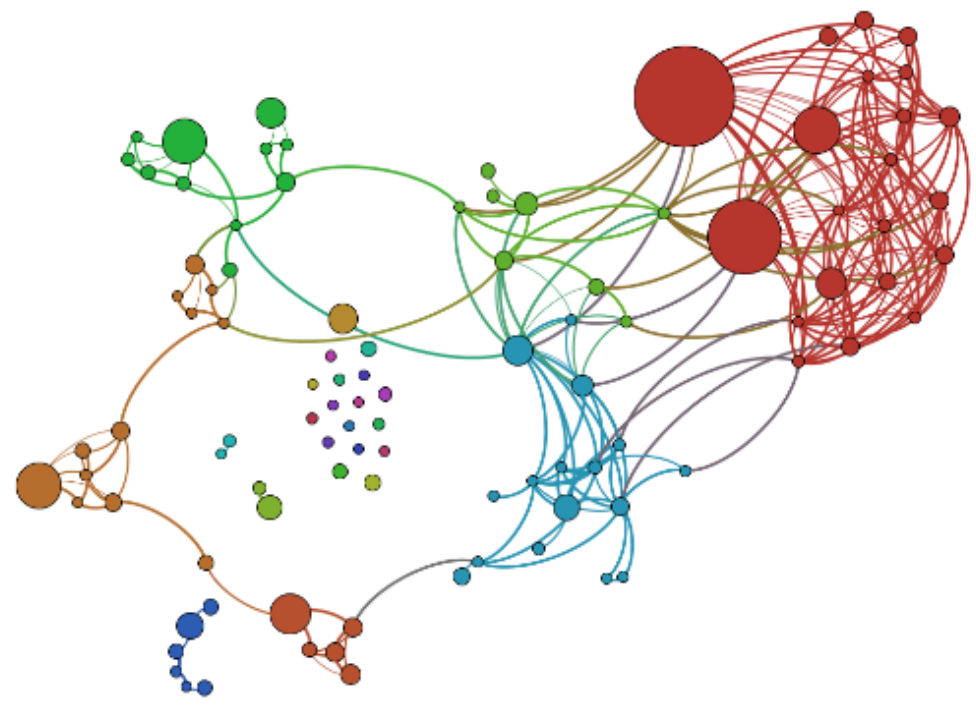

**Figure 6**

To our surprise, like the social network of human beings, passwords have their own "community" and "social network". To make our observation convincing and solid, we further visualize other data set available.

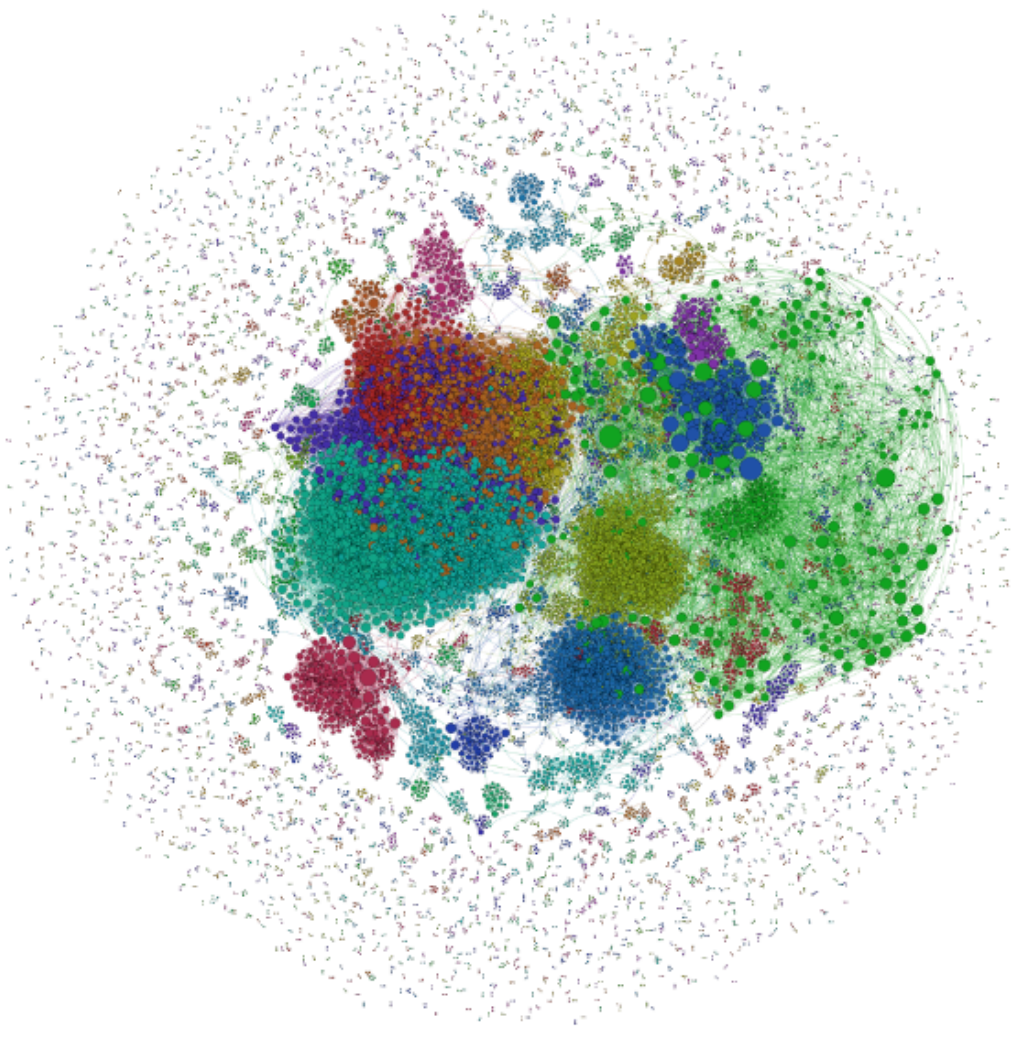

**Figure 7**

Figure 7 the graph of the full 12306 password data after clustering. For the sake of clarity, we filter the isolated nodes in the graph and present the reorganized graph including the nodes that have at least one connection with others in figure 8. It is obvious that the isolated nodes is a single community when analyzing and there is no much point of those single "island".

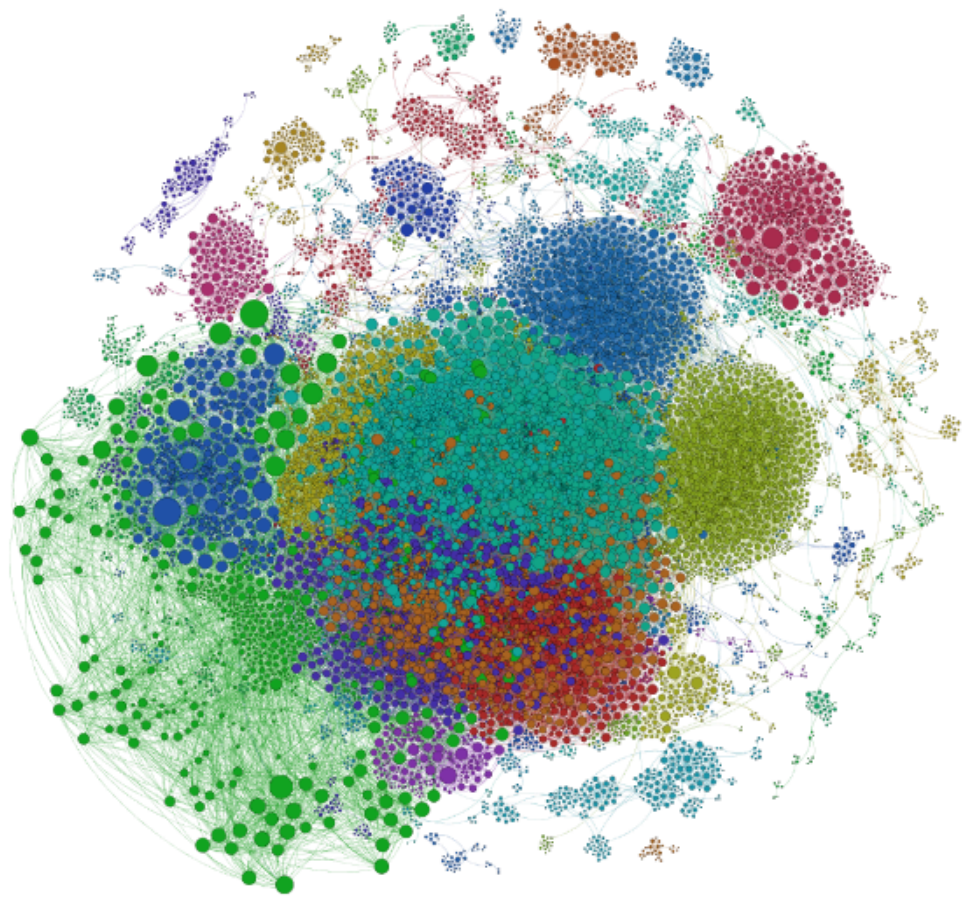

**Figure 8**

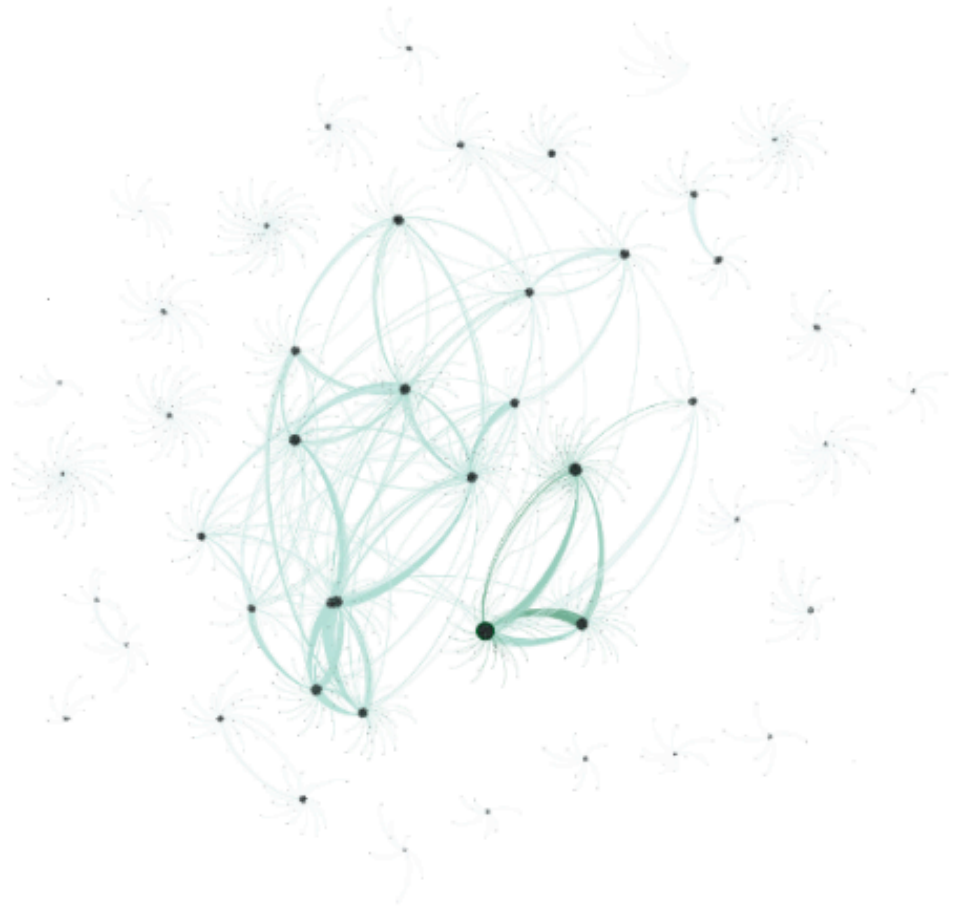

**Figure 9**

Figure 9 is the graph of the Yahoo! data set. To better display the community distribution, we filter the communities whose size is less than 0.1% and the graph is shown in figure 10.

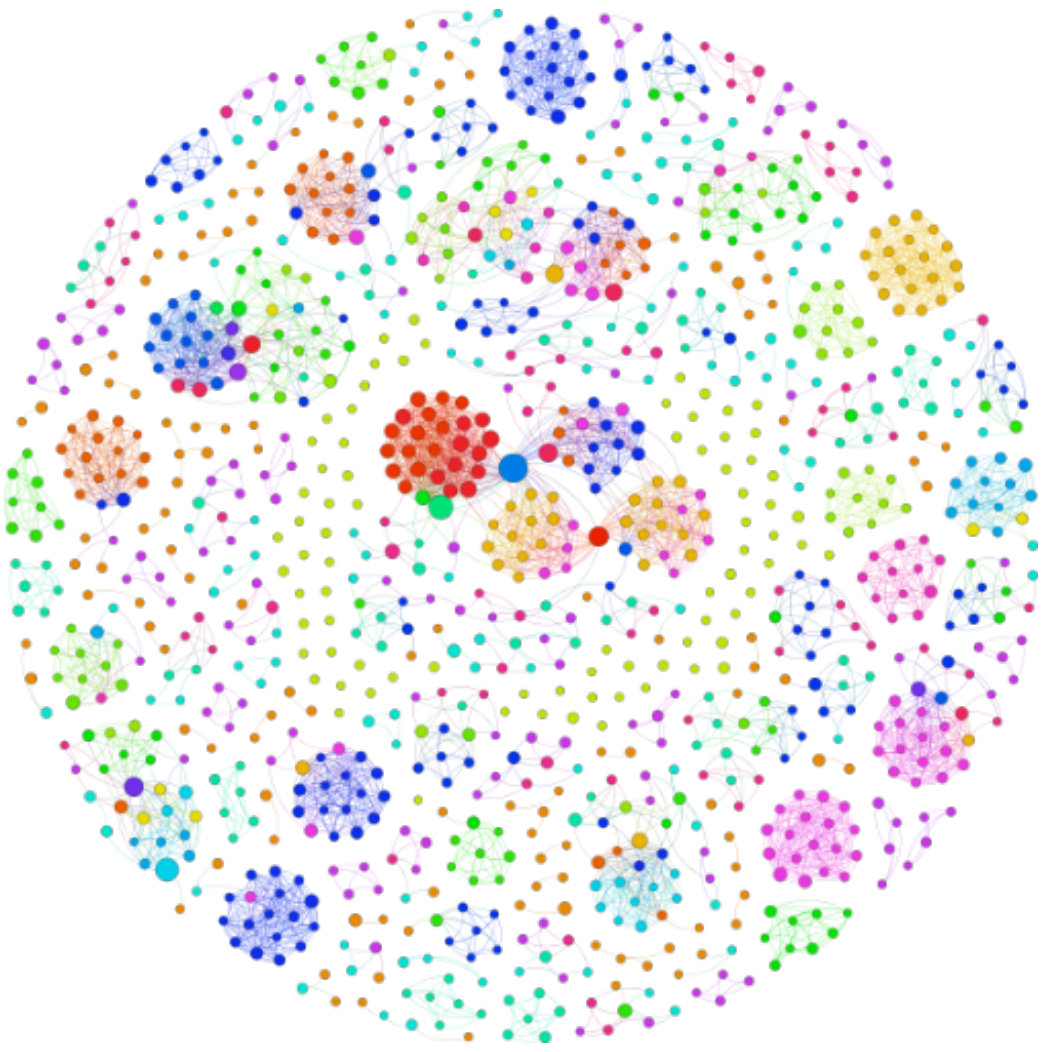

**Figure 10**

As shown in the graph above, the distribution of passwords tends to form communities and clusters. To put it another way, some passwords are closer to other passwords and the whole data set is split into different parts.

After visualizing the data sets mentioned above, we further explore the data. Since the data set form a network and the focus is the interconnection of nodes, the degree of nodes, which is the number of nodes that is connected to the node, represent more information. The Yahoo data is taken as an example and the degree-rank curve is shown in figure 11.

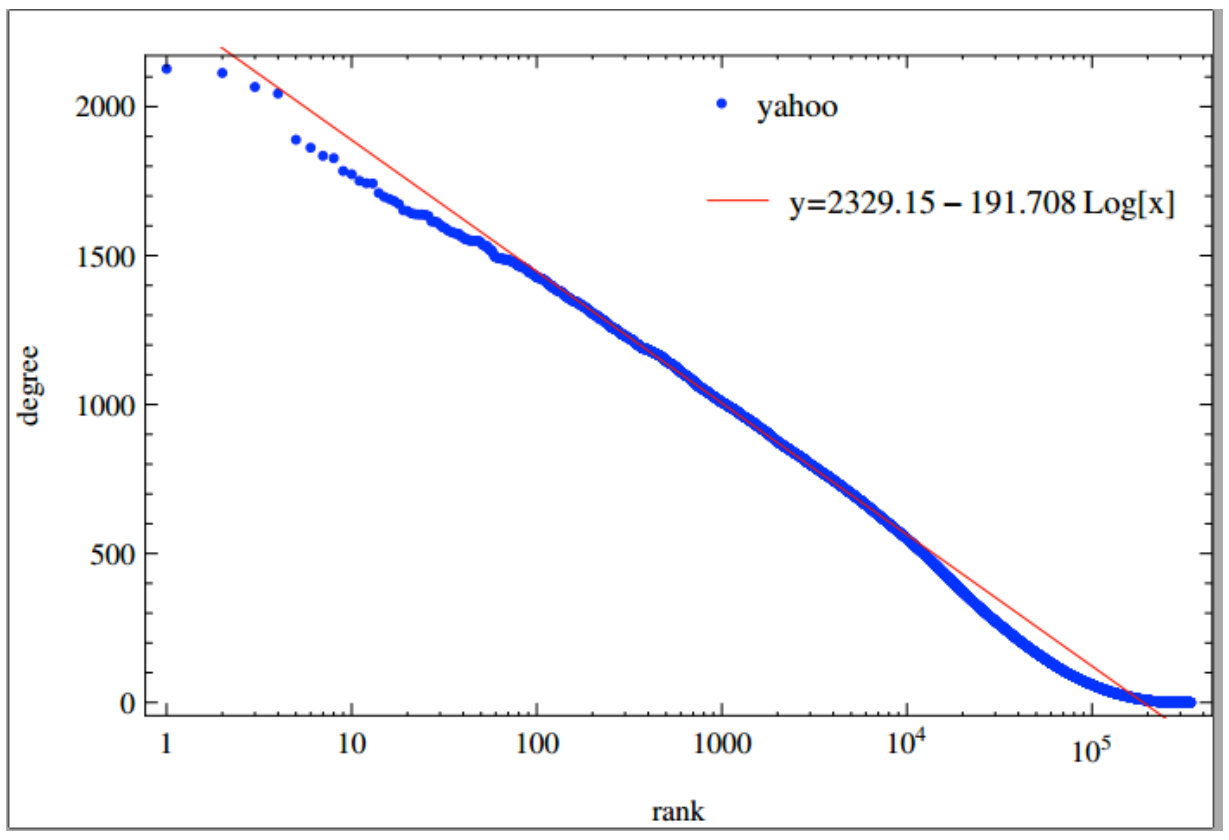

*1) Scale-free network of empirical password sets*

Figure 11 From figure 11, the degree distribution of the Yahoo data follows a power law, i.e.

$$R(k) \sim k\text{^}(-r)$$

So the topological distribution of the password sets is a scale-free network [11]. Results of other data sets show similar conclusion. Until now, we have build a network of the data set and prove it to be a scale free network. The result match with our common sense that popular password is widely used and a great number of users tend to use at least similar password.

In conclusion, the security of one single account and the whole system are not isolated. As the stars in the sky, though they seems to be distributed randomly, through the estimation of their distance (actually the gravity between each other), the ancient human beings divided them into different galaxy and made beautiful stories. In this paper, we might do exactly the same thing and shed light on the “social” network of passwords. From previous research work, we already realize that individuals tend to choose same passwords. Then in this paper, we reveal the fact that people tend to choose similar passwords in a much more higher chance. If we think about it carefully, it does make sense. Since we are individuals and we choose our passwords independently, if we tend to choose same passwords, then the odds of we choose password that is slightly different from each other is much higher. That is where the network of our password begin.

The study of networks originates in the ancient Graph theory and becoming more and more important in both theoretical research and empirical applications. The electric power grid, the WWW, and the pattern of air traffic between airports are early examples of networks in real life. The boom of Social Network in the last decades have made a big step forward in the understanding of social science. The networks of movie actors, scientific collaboration.

Unlike the components like people, airports, routers on the Internet that consist of networks, password is chosen independently and is supposed to be personal and private. We make friends with others and our friends has friends of their own, so the social network is generated. The planes fly between airports and vehicles go along the roads between cities, so we have the traffic and transportation network in geology. The routers on the internet deliver packages between websites and user clients, so we have the so-called Internet. Networks are everywhere. As far as we are concerned, the networks that has been studied so far are PUBIC in some degree. The purpose of the network is to share or transport something, like information, grocery and even virus. As the key to access resources or accounts, password is supposed to be private in the first place. Unfortunately, it turns out that the passwords generate a network that we have never imagine and expose inevitable threaten for numerous accounts and organizations.

*2) Frequency Vs. Degree*

When we only see the discrete list of passwords, frequency is the only factor we could consider to estimate the popularity of a password. Now that we have the network of the whole password set, the concept of Degree Centrality is to be introduced for a better understanding of the password itself.

In Graph Theory, the degree of a node in a graph is the number of edges incident to the node, with loops counted twice [12].

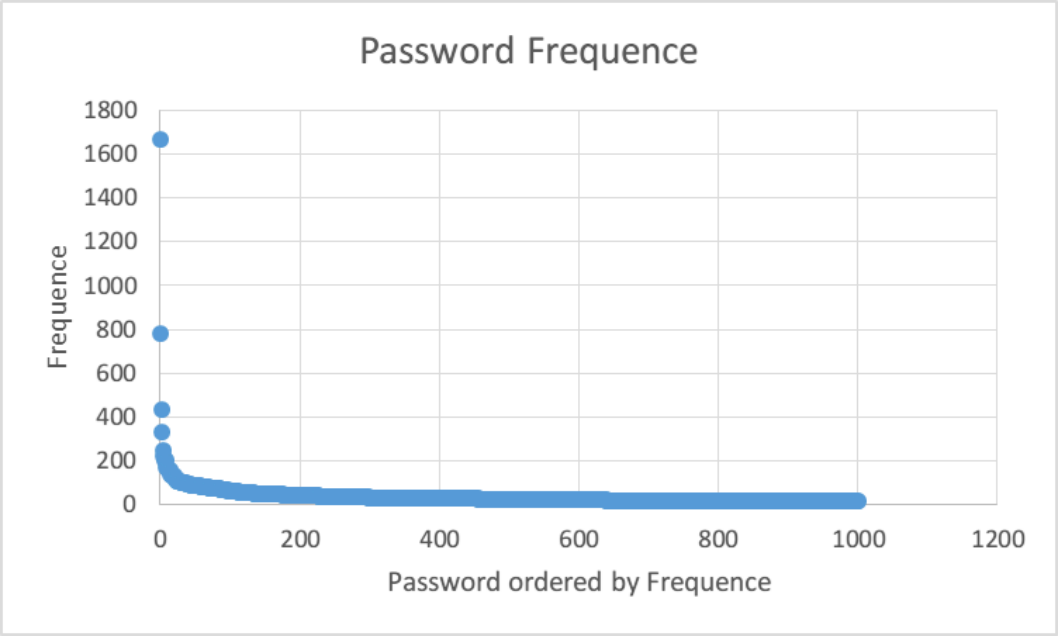

Figure 12

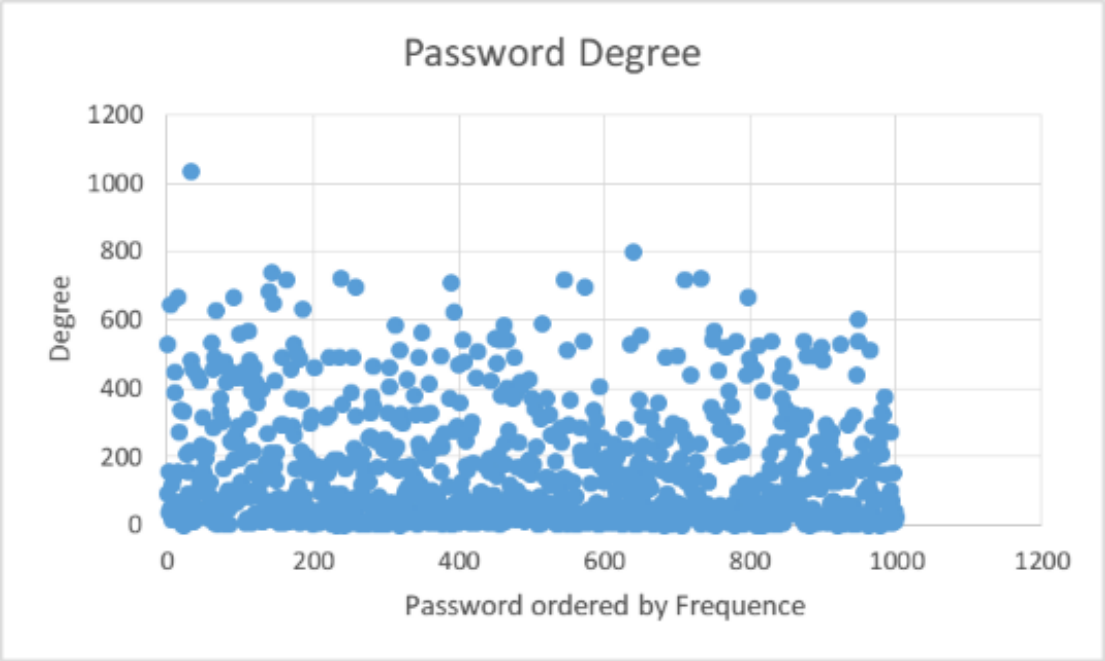

Figure 13

Figure 12 is curve of frequency and figure 13 the degree distribution in the descending order of frequency. As show in figure 13, there is no positive relevance between the degree and the frequency of a password. In other words, higher frequency does not mean greater popularity in the password set [1].

Compared to the Frequency of passwords, in the network of password set, we believe that the node (i.e. the password it represents) with higher degree has more power or impact on the whole password set. According to the definition of degree of a node, a node with higher degree covers more nodes than the other nodes, including the nodes with higher frequency. In a brute-force statistical guessing attacking, this mean the dictionary based on passwords in the decreasing order of degree could compromise more accounts than the dictionary based on passwords in the decreasing order of frequency.

In order to prove the validation and solidness of our inference, we dig deeper into the graph we have built.

## IV. THE STATISTICAL GUESSING MODEL

Assume we have two dictionaries with same number of passwords n:

One is the dictionary Dic-1 based on passwords in the decreasing order of frequency, which including password $p_{11}, p_{12}, p_{13}, p_{14}, p_{15} \ldots\ldots, p_{1(n\text{-}1)}, p_{1n}$. At the same time, the frequency of the password $p_{ij}$ is denoted as $f(p_{ij})$ and

$$f(p_{11}) > f(p_{12}) > f(p_{13}) > f(p_{14}) > f(p_{15}) > \ldots\ldots > f(p_{1(n\text{-}1)}) > f(p_{1n})$$

The other dictionary Dic-2 based on passwords in the decreasing order of degree, which including password $p_{21}, p_{22}, p_{23}, p_{24}, p_{25} \ldots\ldots, p_{2(n\text{-}1)}, p_{2n}$. At the same time, the degree of the password $p_{ij}$ is denoted as $d(p_{ij})$ and

$$d(p_{21}) > d(p_{22}) > d(p_{23}) > d(p_{24}) > d(p_{25}) > \ldots\ldots > d(p_{2(n\text{-}1)}) > d(p_{2n})$$

Then we perform a simulation a brute-force statistical guessing attack based on Dic-1 and Dic-2 separately. Since we we have built a whole graph of the targeted password set, it provide a perfect method to estimate the maximum success ratio.

To estimate the number of potential maximum successful guesses, the concept of neighborhood is introduced.

In Graph Theory, the neighborhood of a vertex v in a graph G(V,E) is the sets of adjacent vertices of G consisting of all vertices adjacent to v.

The neighborhood is often denoted $N_G(v)$ or (when the graph is unambiguous) N(v). Note that the concept of neighborhood we discuss in this paper is the closed neighborhood, in which v itself is included. There is another version of neighborhood is called open neighborhood when v itself is not included [13].

The concept of neighborhood of one vertex can be naturally extended to a set of vertices S, which is the union of all the neighborhoods of the vertices in set S, i.e. the set of all vertices in the original graph which is adjacent to at least one member of S.

Denoted as $N(S)$ and we have

$$N(S) = \bigcup_{i=1}^{\#s} N(v_i),$$

in which #S is the number of vertices in S and is the i-th vertex in S.

Then for two attackers with dictionary Dic-1 and Dic-2 separately, the maximum number of vertices they could cover is

$$N(Dic\text{-}i) = \bigcup_{j=1}^{n} N(p_{ij}),$$ in which i=1 or 2.

Thus their corresponding maximum successful guesses are

$$G_{max}(Dic\text{-}i) = \sum_{p_x \in N(Dic\text{-}i)} f(p_x).$$

To make a further comparison between the difference of frequency and degree, we compute the $G_{max}(Dic\text{-}1)$ and $G_{max}(Dic\text{-}2)$ of dictionary Dic-1 and Dic-2 whose size range from 1 to the size of the original password set. We take the data set of Yahoo as example and the result is presented in figure 14.

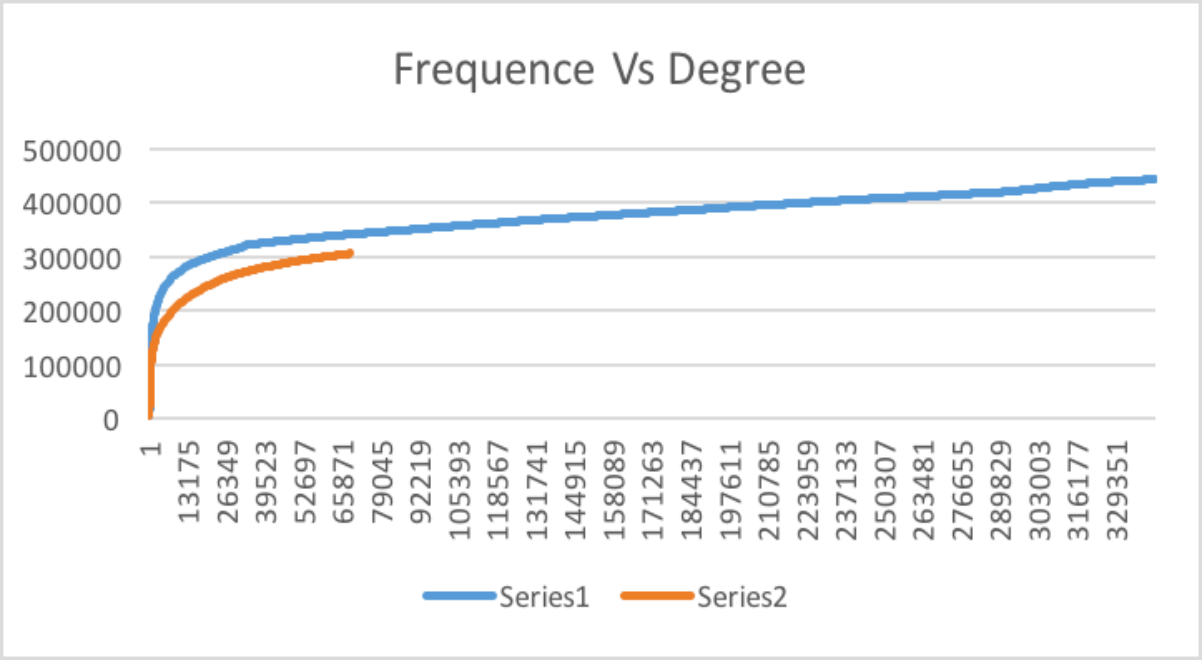

Figure 14

The successful guesses curve also support the effectiveness of statistical guessing based on the dictionary of frequency. The diminishing return, which has also been observed in [2], of dictionary based attacking can also be observed in both of the two curve.

While contrary to our prediction, the dictionary ranked in frequency is more efficient in cracking yahoo, for which we believe is the unbalanced proportion of frequency and degree.

The upper bound of the success ratio using the dictionary Dic-1 and Dic-2 separately is the result of reorganizing the dictionary order to combine the frequency of the node and its neighbor. Of course the attacker can start a new round by searching the closure of the password that has been compromised.

*1) The optimal cracking problem*

In conventional password cracking, the size of dictionary has a significant impact on the success rate of the cracking. In order to achieve higher success rate, most researchers tend to use dictionary as large as possible. The reason is straightforward-----the larger dictionary means the higher possibility of covering more passwords in the targeted set. But due to the efficiency of time and space, all results see diminishing return as the dictionary size grows [14]. The success guessing curves have been observed in almost every previous attempts to crack as more accounts as possible.

Daniel V. Klein [15] made the first attempt to identify the higher efficient sub-dictionary. J Bonneau define a success rate α when introducing α-guesswork to evaluate the number of guesses of an attacker. And Diogo Mónica, Carlos Ribeiro [16] discussed the compression ratio in a implementation of SOM(Self-Organizing Maps) which preserve the topological position passwords.

Since we have built a graph of the password set, the sub-dictionary searching work become easier. Our goal is to find a subset of strings to cover as more passwords as possible. Considering we are dealing with this problem on a graph, if we rephrase the problem a little bit, the goal is to find a subset of nodes that has connections to all the other nodes. That is exactly the definition of Domainating Set in Graph Theory:

**Definition**: Given a graph G = (V, E) with V = {1, 2, ..., n}, a **dominating set** for a graph G is a subset D of V such that every vertex not in D is adjacent to at least one member of D. The **domination number** γ(G) is the number of vertices in a smallest dominating set for G. (a detailed version definition and reference material can be found in [17]).

Hence the minimum size of the dictionary to cover the targeted password set, i.e. its lower bound, equals γ(G). For any password set S={p1,p2,p3,…..pn}, we can construct a graph G=(V,E) through the steps in section 2, then we have transformation f and g:

$S \xrightarrow{f} G$ :for each password in S, add a corresponding vertex in graph G; for every two of the password in S, a edge is attached between them in G.

$G \xrightarrow{g} S$ : traversal of the graph G and the full set of S is obtained.

The complexity of transformation g is linear to n while transformation f has the extra cost of computing the edges, which is $\binom{2}{n} = \frac{n(n-1)}{2}$. Both are polynomial-time. It is legitimate to conclude a exactly one-to-one correspondence between the password in S and the vertex in G. Those transformations show that a algorithm for the minimum dictionary size also provide an algorithm for the minimum dominating set and vice versa. In other words, this two issues are equivalent in terms of computing complexity.

The minimum dominating set problem is a well-known NP-hard problem, which is proved by Garey & Johnson in 1979 [18] and Kann provide a pair of polynomial-time L-reductions between the minimum dominating set problem and the set cover problem, the decision version of set covering was one of Karp's 21 NP-complete problems, in 1992 [19].

Moreover, the minimum domination set is a classic problem in Graph Theory and has been studied for years. For a vertex of degree k, it domains itself and k other vertices [20]. Every n-vertex graph with minimum degree k has a dominating set of size at most $n\frac{1+\ln(k+1)}{k+1}$ ( Arnautov 1974 ,Payan 1975). Exact bound remains to be explored when k is small and McCuaig-Shepherd 1989 and Reed 1996 gives a tighter bound for specified graph. Note that this conclusion also applies to other variants of dictionary based cracking metrics when the corresponding method of building the graph.

## V. CONDLUSION

In this paper, we produce more readable and nifty-looking presentation of empirical password sets in the form of networks. The spatial structure of the password sets is discussed for the first time and is proved to be a scale-free network whose degree follow a power law.

Furthermore, we proposed a statistical guessing model that explain the diminishing return, also known as the cracking curve, that has been observed in nearly every statistical guessing attack. We also prove the equality of the minimal dictionary size problem and the dominating set problem in computing complexity, hence the optimal dictionary size problem is also NP-complete.

## ACKNOWLEDGMENT

This work would not be possible without the inspiration of Zhu Chen, Huang Xin and Wang Ding during our weekly beneficial talks. Thanks my girlfriend Teng Wenying and my family for the generous love and company as always. This work is support by the Peking University Education Foundation (URTP2014PKU003).

## REFERENCES

[1] Schechter, S., Herley, C., & Mitzenmacher, M. (2010). Popularity is everything A new approach to protecting passwords from statistical-guessing attacks. Citeseer, 1–8.

[2] Bonneau, Joseph. "The science of guessing: analyzing an anonymized corpus of 70 million passwords." Security and Privacy (SP), 2012 IEEE Symposium on. IEEE, 2012.

[3] Morris, Robert, and Ken Thompson. "Password security: A case history."Communications of the ACM 22.11 (1979): 594-597.

[4] Klabunde, Ralf. "Daniel Jurafsky/James H. Martin, Speech and Language Processing." Zeitschrift für Sprachwissenschaft 21.1 (2002): 134-135.

[5] Levenshtein, Vladimir I. "Binary codes capable of correcting deletions, insertions, and reversals." Soviet physics doklady. Vol. 10. No. 8. 1966.

[6] Navarro, Gonzalo. "A guided tour to approximate string matching." ACM computing surveys (CSUR) 33.1 (2001): 31-88.

[7] Kukich, Karen. "Techniques for automatically correcting words in text." ACM Computing Surveys (CSUR) 24.4 (1992): 377-439.

[8] Wagner, Robert A., and Michael J. Fischer. "The string-to-string correction problem." Journal of the ACM (JACM) 21.1 (1974): 168-173.

[9] Veras, Rafael, Julie Thorpe, and Christopher Collins. "Visualizing semantics in passwords: The role of dates." Proceedings of the Ninth International Symposium on Visualization for Cyber Security. ACM, 2012.

[10] Boytsov, Leonid. "Indexing methods for approximate dictionary searching: Comparative analysis." Journal of Experimental Algorithmics (JEA) 16 (2011): 1-1.

[11] Barabási, Albert-László, Réka Albert, and Hawoong Jeong. "Scale-free characteristics of random networks: the topology of the world-wide web."Physica A: Statistical Mechanics and its Applications 281.1 (2000): 69-77.

[12] Diestel, Reinhard. "Graduate texts in mathematics: Graph theory." (2000).

[13] Hell, Pavol. "Graphs with given neighborhoods I." Proc. Colloque, Inter. CNRS, Orsay. 1976.

[14] Dell'Amico, Matteo, Pietro Michiardi, and Yves Roudier. "Password strength: An empirical analysis." INFOCOM, 2010 Proceedings IEEE. IEEE, 2010.

[15] Klein, Daniel V. "Foiling the cracker: A survey of, and improvements to, password security." Proceedings of the 2nd USENIX Security Workshop. 1990.

[16] Mónica, Diogo, and Carlos Ribeiro. "Local Password Validation Using Self-Organizing Maps." Computer Security-ESORICS 2014. Springer International Publishing, 2014. 94-111.

[17] Hedetniemi, Stephen T., and Renu C. Laskar. "Bibliography on domination in graphs and some basic definitions of domination parameters." Annals of Discrete Mathematics 48 (1991): 257-277.

[18] Garey, Michael R., and David S. Johnson. "Computers and intractability: a guide to NP-completeness." (1979).

[19] Kann, Viggo. On the approximability of NP-complete optimization problems. Diss. Royal Institute of Technology, 1992.

[20] West, Douglas Brent. Introduction to graph theory. Vol. 2. Upper Saddle River: Prentice hall, 2001.